# Integration and mining of malaria molecular, functional and pharmacological data: how far are we from a chemogenomic knowledge space?

Lyn-Marie Birkholtz[1], Olivier Bastien[2], Gordon Wells[3], Delphine Grando[2], Fourie Joubert[3], Vinod Kasam[4], Marc Zimmermann[5], Philippe Ortet[6], Nicolas Jacq[4], Sylvaine Roy[7], Martin Hoffmann-Apitius[5], Vincent Breton[4], Abraham I. Louw[1§], Eric Maréchal[2§]

[1]Department of Biochemistry and African Centre for Gene Technologies, Faculty of Natural and Agricultural Sciences, University of Pretoria, 0002, Pretoria, South Africa
[2]UMR 5168 CNRS-CEA-INRA-Université Joseph Fourier, Département Réponse et Dynamique Cellulaires; CEA Grenoble, 17 rue des Martyrs, 38054, Grenoble Cedex 09, France
[3]Bioinformatics and Computational Biology Unit, Faculty of Natural and Agricultural Sciences, University of Pretoria, 0002 Pretoria, South Africa
[4]Laboratoire de Physique Corpusculaire de Clermont-Ferrand, CNRS-IN2P3, Campus des Cézeaux, 63177 Aubière Cedex, France
[5]Department of Bioinformatics, Fraunhofer Institute for Algorithms and Scientific Computing, Schloss Birlinghoven, 53754 Sankt Augustin, Germany
[6]Département d'Ecophysiologie Végétale et de Microbiologie; CEA Cadarache, 13108 Saint Paul-lez-Durance, France
[7]Laboratoire de Biologie, Informatique et Mathématiques; Département Réponse et Dynamique Cellulaires; CEA Grenoble, 17 rue des Martyrs, F-38054, Grenoble cedex 09, France

[§]Corresponding authors

Email addresses:

LMB: lynmarie.birkholtz@up.ac.za
OB: ol.bastien@wanadoo.fr
GW: gordon.wells@gmail.com
DG: dgrando@cea.fr
FJ: fourie.joubert@bioagric.up.ac.za
VK: kasam@clermont.in2p3.fr
MZ: marc.zimmermann@scai.fraunhofer.de
PO: portet@cea.fr
NJ: jacq@clermont.in2p3.fr
SR: sroy@cea.fr
MHA: martin.hofmann-apitius@scai.fhg.de
VB: breton@clermont.in2p3.fr
AIL: braam.louw@bioagric.up.ac.za
EM: emarechal@cea.fr

## Abstract

The organization and mining of malaria genomic and post-genomic data is highly motivated by the necessity to predict and characterize new biological targets and new drugs. Biological targets are sought in a biological space designed from the genomic data from *Plasmodium falciparum*, but using also the millions of genomic data from other species. Drug candidates are sought in a chemical space containing the millions of small molecules stored in public and private chemolibraries. Data management should therefore be as reliable and versatile as possible. In this context, we examined five aspects of the organization and mining of malaria genomic and post-genomic data: 1) the comparison of protein sequences including compositionally atypical malaria sequences, 2) the high throughput reconstruction of molecular phylogenies, 3) the representation of biological processes particularly metabolic pathways, 4) the versatile methods to integrate genomic data, biological representations and functional profiling obtained from X-omic experiments after drug treatments and 5) the determination and prediction of protein structures and their molecular docking with drug candidate structures. Progresses toward a grid-enabled chemogenomic knowledge space are discussed.Text for this section of the abstract…

## Background

Malaria is a life-threatening disease affecting half a billion humans in underdeveloped and developing countries. Its global heartland is Africa, with an appalling death toll of 1 to 2 million people every year [1]. Endemic malaria ranges from a permanent incidence in sub-Saharan and equatorial Africa, to a seasonal but recently escalating prevalence in Southern Africa [2]. Four species of malaria parasites can infect humans *via* mosquito transmission: *Plasmodium falciparum* (the species that causes the greatest incidence of illness and death) as well as *P. vivax*, *P. ovale*, and *P. malariae*. They belong to the Apicomplexa phylum, which contains thousands of other parasitic protists of medical and veterinary importance, including plasmodia that infect non-human vertebrates, and other genera, such as *Toxoplasma*, *Neospora*, *Babesia*, *Cryptosporidium* etc [3].

Malaria was eradicated from temperate regions following concerted preventative sanitary actions and after important insecticide campaigns and systematic treatments with available drugs, *i.e.* quinine and chloroquine [4-6]. The last indigenous case was reported in the Netherlands in 1961 [7]. Unfortunately, the prophylactic programs of the 1950's and 1960's, essentially based on insecticide and drug treatments, failed to control malaria in subtropical areas [8]. Resistance to chloroquine spread rapidly [9, 10]. Subsequent attempts to achieve progress in malaria prophylaxis have been characterized by the failure of vaccine development, withdrawal of some insecticides because of toxicity and negative environmental impact, the alarming spreading of mosquitoes' resistance to insecticides and of *Plasmodium* to the very few drugs that have been developed [10-12]. The promise of an effective vaccine is as distant as ever [13]. Current efforts focus on chemoprophylaxis using artemisinin, an antiplasmodial molecule from *Artemisia annua*, and derivatives which can be produced efficiently and cheaply. However the scientific community is worried that plans for the extensive use of artemisinin might be ruined by emergence of the parasitic resistance it will almost certainly trigger, sooner or later [14, 15]. Given the small number of available drugs and the resistance they have already induced, discovery of new targets and of new drugs remains a key priority.

A major landmark in the history of malaria was the launch of a collaborative genomic sequencing program in 1996 ([16, 17]; for review: [18-21]). In November 2002, the complete genome of the 3D7 strain of *P. falciparum* [22] and a whole genome-shotgun of *P. yoelii yoelii* [23] were released, followed by whole-genome shotguns of *P. berghei*, *P. chabaudi*, and *P. vivax*, and genomic sequencing of *P. gallinaceum*, *P. knowlesi* and *P. reichenowi* which are still under progress [21, 24]. This unprecedented effort to sequence genomes of eukaryotic pathogens was a technical challenge, because the extreme compositional bias of *Plasmodium* DNA (>80% A+T in *P. falciparum*) accounted for the instability of genomic fragments in bacteria [17-19] and complicated assembly of contigs [19]. Among eukaryotes, the *Plasmodium* genus is therefore the best documented at the genomic sequence level, with well established syntenic relations. At the level of the Apicomplexa phylum, additional complete genomes of *Cryptosporidium*, *Theileria* and *Toxoplasma* have been either released or announced [25-26].

All *Plasmodium* molecular data have been collected and organized in the PlasmoDB public database as early as sequencing outputs were made available [27-30]. The architecture of the relational database was designed following biologically relevant relationships, *i.e.* the "gene to mRNA to protein" dogma, using the Gemomics Unified Schema [28-29], and ensures that gene loci are linked to annotation using the Gene Ontology standards [30]. A genome browser allows navigation along chromosome sequences and the viewing of multiple *Plasmodium* species at one glance, based on syntenies. Predictions of protein domains, post-translational modifications, subcellular targeting sequences, etc. are included. Furthermore, PlasmoDB is currently the only site where molecular data are (1) tentatively clustered based on homology, (2) linked to generic schemes designed to view metabolic pathways, and (3) linked to X-omic functional information (transcriptome, proteome, interactome): any biologist can exploit these integrated data with basic or combined queries [27-30]. It is therefore the first resource designed to help the scientific community to turn genomic data hopefully into gold, *i.e.* appropriate biological knowledge that can accelerate the design and introduction of new therapeutic strategies. PlasmoDB operates inside ApiDB, a master web portal for apicomplexan genomes [29, 32].

Contrasting with this clever, integrated and user-friendly access to molecular and functional data, the proportion of genes for which a biological function has been inferred appears like a curse. Only 34 % of the *P. falciparum* genes could be assigned a function, based on detected sequence homology with characterized genes from other organisms [22, 33, 34]. The most-sequenced eukaryotic group appears therefore as the worst functionally annotated. The first version of the *P. falciparum* genome was estimated to code for 5268 proteins, out of which 3208 did not have any significant similarity to proteins in other organisms to justify provision of functional assignment. This proportion of uncharacterized genes was further increased by 257 additional sequences, which had significant similarity to proteins, described as hypothetical, in other organisms [22]. The 2005 updated EMBL version of the 3D7 complete genome (generated at Sanger Institute, The Institute for Genomic Research, and Stanford University; version 2.1) was still predicted to encode as many as 3548 hypothetical proteins (65.6 % of total). This figure is the worst ever recorded for a eukaryotic genome (Table 1) and a clear limitation to any *in silico* exploration of the malaria biology.

<Table 1>

Since absence of evidence is not evidence of absence, the scientific community faces a serious epistemological problem when trying to derive conclusions from a small genome (the number of genes is similar to that of yeast), in which two thirds cannot be used in meaningful analyses. Sustained effort is therefore required to improve functional annotation methods, and contribute to the PlasmoDB gene descriptions. Different teams in the world have attempted to address this problem. In this paper, we provide an overview of some of the theoretical and practical developments that were introduced to improve detection of similarities between *Plasmodium* sequences and distantly related organisms.

A second difficulty also related to sequence homology detection is the reconstruction of molecular phylogenies for malarial genes. Accurate molecular phylogenies are particularly important since the *Plasmodium* cellular organization is the result of multiple endosymbioses involving an ancestral alga [3, 35, 36], at the origin of a plastid relic, *i.e.* the apicoplast ([37, 38]; for review: [39-43]). Comparative phylogenomics focusing on malarial plant-like genes proved to be a valid strategy to detect potential targets for herbicides that act as antiplasmodial [44]. *In silico* analyses combining molecular phylogeny and targeting sequence prediction allowed a first rationalized mining of the apicoplast function [45]. The identification of all biological processes that have been inherited through lateral gene transfers from the ancestral alga (the algal sub-genome) is therefore one of the most important outputs one expects from comparative phylogenomics. Although molecular phylogenies of a few genes can be achieved with conventional methods, combined with expert visual analysis, it is difficult to carry out high-throughput phylogenetic determination at a genomic scale. We summarize how the question of automatic assessment of orthologies has been addressed, particularly with the OrthoMCL [46] and TULIP [34, 47] approaches.

A third difficulty is the representation of knowledge of malaria parasite biology. Description of gene function follows the guidelines of the Gene Ontology (GO) structured vocabulary [31]. GO is a standard adopted by all the scientific community to circumvent problems raised by heterogeneous key word annotations. GO also complies with *in silico* management (and mining) of information. Genes coding for enzymes can further be linked to metabolic scheme(s) designed using generic methods, *i.e.* KEGG [48] or MetaCyc [49], or specifically designed for *Plasmodium*, *i.e.* the Malaria Parasite Metabolic Pathways (MPMP) [50, 51]. Pathways based on generic methods do not include representations for cell compartmentalization. This information, which has been included in the MPMP, is essential to understand the metabolism of water soluble intermediates inside and between different cell compartments, and is absolutely critical for pathways involving lipophilic compounds localized in disconnected membranes. Eventually, knowledge should be represented in a way that is usable for the organization and analysis of functional data. We discuss here how each approach complies with these theoretical and practical constraints.

Fourthly, the difficulty of organizing molecular data into knowledge representations becomes more pronounced in the analysis of global datasets arising from X-omic (transcriptome, proteome, interactome) experiments. We examined the strategies, methods and tools that have been used or designed in order to link malarial molecular and functional data in the most versatile way. In particular we describe the MADIBA tool developed for local analyses of transcriptomic outputs.

A fifth difficulty, once a target gene has been identified for possible antimalarial intervention, is the process of reaching a decision on the entry into costly drug- or vaccine-development programs. Besides experimental validation criteria, *in silico* experiments can be a useful tool to help characterize a possible new target. These include determination of protein three-dimensional structure and determination of possible binding ligands through *in silico* protein-ligand docking. To date, the structures of less than 70 malaria proteins have been determined and made available *via* the Protein Data Bank [52]. We argue that a repository for all known malaria resolved protein structures and structural models should be initiated, curated and maintained. No prediction of protein druggability has been investigated. Access to such repository might be invaluable for drug discovery projects: virtual screening of hundreds of thousands of potential drugs, making use of protein structures of a whole family has been achieved using computer grid resources with the WISDOM I project [53], and oncoming WISDOM II.

This review does not pretend to provide any complete panorama on malaria molecular, functional and therapeutical genomics or to introduce panaceas. We focused on the important difficulties listed here for global analyses and high throughput approaches as five major challenges for the future, *i.e.*, 1) comparisons of the compositionally atypical malaria gene and protein sequences, 2) high throughput molecular phylogeny assessment, 3) usable and interoperable representations of metabolic and other biological process, 4) versatile and local integration of molecular and functional data obtained from X-omic experiments, and 5) determination and prediction of protein structure and subsequent virtual ligand screening on candidate therapeutical targets. Linking protein structures with ligand structures is a pivotal step for a chemogenomics knowledge base, in which functional and structural knowledge deriving from malaria genomics and post-genomics might be connected with the space of small molecules containing known and potential drugs.

# Integration of malaria genomic, post-genomic data and chemical information: current status and future challenges

### Sequence comparisons of compositionally-biased and insert-containing malaria genes and proteins.

The extreme A+T bias in *Plasmodium* DNA has been a recurrent problem for malarial genomics and post-genomics. It was responsible for instability of genomic segments in *E. coli* and difficult assembly during the sequencing process [19]. It implied debated modifications and combinations of automatic gene detection methods for open reading frame prediction [54, 55]. The nucleotide bias has been demonstrated to be responsible for the protein composition bias [56-58].

Combined with the frequent insertions seen in malarial proteins, the amino acid compositional bias is critical for routine sequence comparison methods, particularly because it can compromise the statistical analyses and sorting of BlastP alignments [58]. Indeed, an alignment algorithm comes with a statistical model implemented in the code, particularly in the Blast package [59], on which users rely to assess the significance of the alignment, and to sort them. It is therefore difficult to discuss the current view on sequence comparison methods and statistics independently. Two major statistical models are used to test alignment scores. The most common test is an estimate of the *E-value* (short for Expectation value), *i.e.* the number of alignments one expects to find in the database by chance, with equivalent or better scores. It can be determined from the complete distribution of scores. The BlastP associated statistics defined by Karlin and Altschul [59] are based on the probability of an observed local alignment score according to an extreme value distribution. The validation of the Karlin-Altschul *E-value* computation model requires two restrictive conditions: first, individual residue distributions for the two sequences should not be 'too dissimilar' and second, sequence lengths 'should grow at roughly equal rates' [59]. Validity restrictions listed here are fully acceptable when dealing with protein sequences of average lengths and amino acid distribution, and BlastP is a good compromise for batch analyses of genomic outputs. However, the compositionally biased proteome of *Plasmodium falciparum* fall outside of the validity domain for a BlastP comparison with unbiased sequences [58, 60]. One of the reasons explaining that 60 % of the *Plasmodium* sequences did not have any apparent homology with sequences from other genomes may not be that most malarial genes are unique in the living world, but that the BlastP semi-automatic annotation procedure was technically limited. Some missed sequences could be retrieved by adding protein structural information (such as hydrophobic cluster analysis, [61]), but these methods require visual expertise and cannot easily be automated. Iterating the BlastP procedure has also proven to be helpful in detecting missed homologies [62], providing evidence for the initial failure of alignment significance detection.

An alternative method to assess the relevance of a pairwise alignment was introduced by Lipman and Pearson [63]. It uses the Monte Carlo techniques to investigate the significance of a given score calculated from the alignment of two real sequences. It can be used to sort results obtained by any comparison methods, including BlastP, although this has not yet been achieved at a massive scale. It is currently used to estimate the probabilities of Smith-Waterman comparisons [64]. The asymptotic law of *Z-value* was shown to be independent of sequence length and amino acid distribution [65] and is therefore fully valid for malaria sequence comparisons. Bastien et al. [60] demonstrated the TULIP theorem (theorem of the upper limit of a score probability) assessing that *Z-values* can be used as a statistical test, a single-linkage clustering criterion and that $1/Z\text{-}value^2$ was an upper limit to the probability of an alignment score whatever the actual probability law was. From the TULIP theorem and corollaries [47, 60], the comparison of a protein to a given reference sequence *a*, weighed by an alignment score, is characterized by a bounded probability that the alignment is obtained by chance. In practice, a *Z-value* table can be analyzed using the TULIP theorem to detect pairs of proteins that are probable homologues following a *Z-value* confidence cutoff. For instance, a *Z-value* above 10 allows an estimate that the alignment is significant with a statistical risk of $1/Z\text{-}value^2$, *i.e.* 0.01. A version of the BlastP algorithm, implemented with *Z-value* statistics, which should be helpful to refine malaria sequence comparisons, is currently beta-tested under the supervision of Olivier Bastien at Grenoble University, and will be made available at the Pretoria University ACGT web portal.

Improvement of automatic annotation procedure are therefore expected, in particular by combining sequence comparisons with GO term associations (*e.g.* GOtcha; [33]) a complementary approach to the annotation based on the combination of GO terms with functional X-omic response patterns (*e.g.* Ontology-based pattern identification - OPI - following the guilt-by-association principle; [66]) (see below).

## Genome-scale assessment of malaria molecular phylogenies

The amino acid compositional bias and high insert content of malaria proteins is also a disturbing factor when attempting to reconstruct phylogenies. Conventional methods used for phylogeny reconstructions based on multiple alignments can be used in conjunction with visual judgment, (*e.g.* [67]), with qualitative decisions on how protein segments "align well". However, such manual pretreatment cannot be undertaken for all known genes. Alternatively, high throughput molecular phylogenies can be derived from massive all-against-all comparisons, based on pairwise alignments [68]. The questions of the statistical accuracy and maintenance of high throughput phylogenetic reconstruction are critical when including compositionally atypical and high insert containing sequences.

The output of an all-by-all comparison of *n* protein sequences is an *n* x *n* table. According to the output table processing, it can be either totally recomputed at each database update, or stored and updated by computing complementary tables. Information is extracted from the output table to help reduce complexity and diversity at the sequence level. Sets of sequences sharing features are named "clusters" [69].

A first massive comparison project, OrthoMCL, was designed to cluster malaria genes based on their sequence similarity with genes of 55 other genomes (> 600,000 sequences), using the BlastP algorithm and Karlin-Altschul *E-value* statistics to build the all-against-all comparison table [46, 70]. As mentioned above, and discussed recently [68] for massive comparisons based on BlastP/*E-values*, *i.e.* COG [71], Tribe [72], ProtoMap [73], ProtNet [74], SIMAP [75] and SYSTERS release 4 [76], there is no theoretical support to justify that an *E-value* table can be converted into a rigorous and stable metric. The handling of the output *n* x *n* table of *E-values* requires pragmatic post-processing normalization, including asymmetric corrections of *E-values* obtained after permutation of the two aligned sequences or consensus *E-value* computation after alignment with different algorithms. The *E-value* table can be converted into a Markov matrix (*e.g.* Tribe, SIMAP, OrthoMCL), or a close graphic equivalent, *i.e.* graphs connecting protein entries with *E-values* as weights for graph edges (*e.g.* COG, ProtoMap, SYSTERS), a representation that has been exploited in the OrthoMCL project. The protein sets are organized either by detecting graphs and sub-graphs following pragmatic rules, with granularities depending on *E-value* thresholds, or by distance clustering using *E-value* as a pseudo-metrics, or by Markov-random-field clustering. None of the organization of the protein sequences obtained through these methods can be named a spatial projection, and none of the obtained clusters can be represented as a phylogenetic reconstruction. Eventually, the economy of computing *E-values* in an all-by-all comparison experiment is lost in the updating process that requires a complete re-calculation. In spite of these drawbacks, massive comparisons based on BlastP/*E-values* have been undertaken because they were less CPU-demanding than other methods. The OrthoMCL clustering method based on BlastP/*E-value* represents therefore a pragmatic reduction of the protein diversity, and phylogeny reconstruction require post treatments within each clusters. OrthoMCL flags probable orthologous pairs identified by BlastP as reciprocal best hits across genomes. Access to OrthoMCL groups is linked to the PlasmoDB GUS underlying database, allowing multiple queries with other PlasmoDB data and information, and allowing additional cross-species / cross-phylum profiling of the BlastP/*E-value*-supported orthologues.

A more CPU-demanding alternative method for massive all-against-all protein sequence comparison uses Smith-Waterman/*Z-values* rather than BlastP/*E-values*. This method has been initiated for the ClusTR protein sequence clustering, underlying the UniProt/Integr8 knowledge base at the European Bioinformatics Institute, EBI [77]. Because of the properties of the *Z-value* statistics detailed above, it is the solution of choice when comparing compositionally biased and high-insert containing sequences. Additionally, for any set of homologous proteins, it is possible to measure a table of pair-wise divergence times and build phylogenetic trees using distance methods [47]. These trees are called TULIP trees. TULIP trees were compared to phylogenetic trees built using conventional methods, for instance the popular PHYLIP [78] or PUZZLE [79] methods based of multiple sequence alignments. TULIP trees proved to perform as well in any unbiased sets of proteins. Moreover, some phylogenetic inconsistencies in trees built with multiple-alignment based methods, particularly including subsets of compositionally biased sequences, or with low bootstrapping values, could be spectacularly solved with the TULIP tree [47]. An advantage of the phylogenetic inference from the CSHP over that obtained from multiple alignments lies precisely in the TULIP tree construction from pair-wise alignments. Whereas the addition or removal of a sequence can deeply alter the multiple alignment result, and the deduced phylogeny, the *Z-value* and divergence time tables that serve to reconstruct the TULIP trees are the result of a Monte Carlo simulation, which is a convergent process at the level of the pair-wise comparison and is not altered by database updates. As a result, whereas a phylogenetic database computed from multiple alignments would require a complete and increasing computation for any update, the TULIP tree calculation simply requires the calculation of the {new}-by-{old} and {new}-by-{new} *Z-values* and divergence times. A mapping of each *Plasmodium* sequence can therefore be obtained and updated following all-against-all pairwise comparisons based on *Z-value* statistics. The CPU-cost required by the Smith-Waterman comparison method and by the Monte-Carlo simulation used to compute *Z-values* will be compensated in the future by implementing *Z-values* on the BlastP heuristics. A high-throughput assessment of molecular phylogenies of *Plasmodium* genes based on BlastP/*Z-value*, including all recorded genes in public databases, will therefore be feasible and upgraded at the pace of public database updates.

User access to molecular phylogenies, which has been designed in the OrthoMCL project in a very practical and user friendly way, is essential to mine the genome for clues to therapeutic opportunities. The most obvious approach is to detect protein sequences that are excluded or diverge strongly from the mammalian proteome. More subtly, the question of the plant/algal sub-genome of Apicomplexans has been demonstrated to be a source of therapeutic targets for herbicidal drugs (*e.g.* apicoplast lipid - fatty acid, isoprenoid -syntheses, plant-like targets localized outside plastids - folate metabolism, tubulin -, etc.). Criteria for confirmation of a plant/algal sub-genome in *P. falciparum* include molecular similarities with plant genes, and will therefore benefit from future progress in high-throughput molecular phylogenies. Information on sequence and sequence similarity are not sufficient to highlight functions that are, for instance, unique to plants: they have to be linked to appropriate knowledge representation of the biological function of each gene and each process in which gene products play their roles.

## Knowledge representations of the biological function

Having in hand a table of gene entries, with summarized annotations is not sufficient to handle genomic information. Data can be organized based on biological principles so as to reflect our current knowledge at best, and to be viewed at a glance in global X-omic experiments, or for comparative purposes. Biological (and chemical) knowledge is primarily produced in the form of research publications, books, patents and other un-structured texts, and since two decades in structured databases (biologists are used to fill Genbank forms prior to paper submissions). The general principles for the *in silico* organization of knowledge are therefore based on more or less rigorous definitions of semantic networks. Data and information can be organized as semantic networks, following combinations of ontologic hierarchies and praxeologic schemes [80]. In brief, an ontological hierarchy is designed to organize entities (here biological or chemical entities) following inclusion/subsumption principles ("A" is part of, is a component of, etc. "B"). Best known examples are the taxonomic trees ("species" is part of "genus", etc.) or the Gene Ontology, or GO [31], although this later was built from a lose definition of the ontology (see below). A praxeologic scheme is designed to organize the activity, the function achieved by an entity (for instance the enzymatic activities caused by proteins, or the biologic effect caused by drugs) based on the transformation or alteration of an entity into another, through time ("C" is converted into "D"). By contrast with ontologies, which should be stable and non-conflicting hierarchies, praxeologies can describe cyclic processes and can vary over time. Best know examples are metabolic graphs and fluxes, and their variations in different physiological conditions. Interestingly, the enzymatic activities can be organized following hierarchical principles (e.g. the EC numbers proposed by the Enzyme Commission) and a praxeologic scheme such as a metabolic pathway is therefore linked to one or more ontologies for the metabolites (hierarchical categorization of molecules) and for enzymes (hierarchy following the GO, the EC, etc.). In practice, although they are of very distinct nature, ontologic hierarchies and praxeologic schemes are handled as semantic networks, called "ontologies" *senso lato*.

Following recommendations of the Gene Ontology (GO) consortium [31], malaria gene function was defined using a semantic network organized in three hierarchical axes, *i.e.* "molecular function", "biological process" and "cellular component", according to a controlled hierarchical vocabulary. The GO-based annotation circumvents the problems raised by heterogeneous key word annotations, allowing subsequently a cross-species comparison with genomes annotated similarly, and complies with *in silico* management requirements. Genes thus annotated can further be embraced in higher order representations of biological knowledge.

Concerning enzymes, entries can be linked to graphical representations of the metabolic reactions they catalyze. To that end, in the same way genes were defined following the GO procedure, enzyme substrates and products are defined following a chemical ontology (CO), and the reaction itself has to be defined following an enzymatic reaction ontology. A reaction can be viewed as a small graph in which the enzyme is associated with the line that connects the nodes corresponding to the substrates and products. Metabolic pathways are connected by shared nodes, and can be viewed at different scales. Current representations of malaria metabolism have been made available using generic methods, *i.e.* KEGG [48, 81] or MetaCyc [49, 82, 83], or specifically designed for *Plasmodium*, *i.e.* the Malaria Parasite Metabolic Pathways (MPMP) coordinated by Hagai Ginsburg at the Hebrew University of Jerusalem [50, 51].

The Kyoto Encyclopedia of Genes and Genomes (KEGG) resource provides a set reference of metabolic schemes, manually designed so as to represent all possible primary metabolic reactions, and formatted with the KEGG Markup Language (KGML) [48]. These reference schemes can be explored on the KEGG web portal, with a very clear view of the global metabolic map, connecting all pathways.

Organism-specific schemes are generated, based on sequence similarities (using the KEGG Orthology, or KO orthologue identifier) with references to the KEGG Gene catalogs. Thus 79 metabolic schemes have been generated for *P. falciparum*. Each scheme provides links to metabolite (substrates, products, co-substrates) information, and enzyme descriptions following the Enzyme Commission (EC) classification. EC numbers give access to multiple sequence alignments, protein motifs, genomic mapping, links to Genbank, UniProt, PDB, etc. Tilling KEGG schemes from different organisms highlights metabolic similarities and differences, and could be of help for anti-malarial purposes, highlighting for instance metabolic reactions occurring in *Plasmodium* and not in humans. However, the maps are designed based on the enzymatic reactions and, for instance, the fatty acid synthesis due to the type 1 fatty acid synthase (FASI, a multiprotein complex) from the human cell cytosol, strictly overlaps with the type 2 fatty acid synthase (FASII dissociated enzymes) from the *Plasmodium* apicoplast. Information on the protein structure and cell compartmentalization of the enzymes would have been sufficient to distinguish between FASI and FASII. Furthermore, metabolites that are generated in one compartment (*i.e.* diacylglycerol generated in one of the numerous cell membranes) may not be available for an apparent downstream reaction occurring in another compartment. Thus, used without caution, the KEGG schemes may seem to be fully valid for the entire living world, with a misleadingly clear and fully cross-connected global overview of metabolism, and can lead to unrealistic representations. The missing enzymes, which have been experimentally assayed, are not shown, and it is unclear whether gaps within pathways are due to absent enzymes or to incomplete data. Additionally, the KEGG representation is not intended for the design of schemes other than those pre-defined. KEGG outputs need therefore to be reexamined for accuracy of interpretation (see below).

As an alternative, the design of MetaCyc schemes for *Plasmodium*, called PlasmoCyc, has been initiated [49, 82]. As for KEGG, a reference of the complete metabolic pathways has been designed manually and loaded in a MetaCyc database. In contrast with the fully connected overview of the KEGG metabolic map that can make the user overconfident, the MetaCyc global view of metabolism is fragmented, reflecting knowledge gaps, incomplete design of some pathways (such as the tricky pathways for lipid syntheses), and the versatility of the MetaCyc tool for implementation of new schemes. Using *P. falciparum* gene annotation and information of the MetaCyc reference database, the PathoLogic module of the Pathway Tool Software [84] allows the generation of a Pathway/Genome Database (PGDB). A total of 113 metabolic pathways (complete or fragmented) have been generated for *P. falciparum*. As in KEGG, each graph gives access to metabolite, reaction and enzyme information. The Gene-Reaction Schematic (GRS) representation allows a visualization of the relation between the genes, the enzymes, the catalyzed reactions, even in the case of complex or multienzymatic proteins. This model is therefore useful to distinguish for instance FASI and FASII proteins. It is further useful for *Plasmodium* proteins that are often multienzymatic. For each reference pathway, the occurrence or absence of enzyme homologues in the *Plasmodium* genome is documented. As in the case of KEGG, compartmentalization information is missing from the pathway model, in spite of some effort to highlight some specific pathways (for instance the apicoplast fatty acid synthesis).

A synthetic representation of *Plasmodium* biological processes can further be viewed at the Malaria Parasite Metabolic Pathways web portal [50]. Among 120 schemes representing numerous cellular processes, half represent metabolic pathways and were fully designed for malaria researchers. All schemes were built using KEGG pathways, cleaned of irrelevant information and curated by international experts. The quality of each representation is therefore very high and specialized, and benefits from a sustained effort in biological and molecular investigation and validation. Missing data are concisely documented. Most importantly, enzyme subcellular localization is shown. Graphs are not automatically generated, but drawn by experts. While MetaCyc graphs are self generated taking into account all information of the underlying database, the MPMP scheme are not self generated: in particular, the subcellular localization representation is not generated automatically from a subcellular attribute filled in the underlying database. Thus, although the MPMP representations are of higher quality, their update is not dynamic and cannot be used for *in silico* graph-based treatments.

From this short overview, it is clear that metabolic representations should be carefully used. On one hand, generic approaches (KEGG, MetaCyc) do not include cell compartmentalization data and the chemical ontology for lipids is not finished. On the other hand, the curated MPMP representations are based on high-quality data but they are static, with no graph management tools, giving click-access to remote information. Future challenge will be to design underlying models for building graphs from genomic data (following the GO), metabolite data (following a stable Chemical Ontology), reaction connections and compartmentalization information for both gene products and metabolites. Construction of PlasmoCyc benefited from information of the MPMP [50, 82], and MPMP was designed after cleaning KEGG schemes. Models for other knowledge bases and graphs should benefit from the important effort of the MPMP in defining metabolic and other biological processes.

## Connecting functional schemes and ontologies with post-genomic global functional analyses

The application of functional genomics strategies to assign functionality to each gene product of an organism has recently attained increased attention in the field of post-genomics research of *Plasmodia*, particularly oriented towards the discovery of new drug leads (therapeutics) and novel drug targets. This includes the understanding of the transcriptome, proteome and interactome of the parasite to elucidate mode-of-action of inhibitory compounds, allow optimization of such inhibitor activities, explain resistance mechanisms to known drugs, chemically validate potential drug targets and ultimately identify and/or functionally describe new drug targets (Figure 1).

The *P. falciparum* transcriptome has been extensively investigated resulting in comprehensive profiles of transcript expression throughout the complete life cycle of the parasite [85, 86]. The overall conclusions demonstrated that the majority (~87%) of the predicted genes are actively transcribed during the lifecycle but that 20% are specific of the intraerythrocytic developmental cycle and are produced in a periodic nature in a ‘just-in-time’ fashion. These early reports have been followed by investigations designed to answer numerous biological questions, including transcription and post-transcription specific aspects of the regulation of protein expression, transcriptional machinery and inheritance, interstrain conservancy, gametocytogenesis and antigenic variation control mechanisms [87-91]. A high degree of correlation exists between the *in vitro* and *in vivo* transcriptomes of *P. falciparum* with an overexpression seen for genes encoding a sexual stage antigen as well as gene families that encode surface proteins, providing interesting new vaccine candidates [92, 93].

Although transcriptomics remains the most comprehensive, highest-throughput methodology to provide vast amounts of data on a global level, it does not specify the dynamic, functional units of a cell. Proteomics studies are essential to conclusively prove mechanistic changes and explain global protein expression profiles, differential protein expression, posttranscriptional control, posttranslational regulation and modifications, alternative splicing and processing, subcellular localization and host-pathogen interactions (Figure 1). Reassuringly, there is a good correlation between the abundance of transcripts and the proteins encoded by these during the *P. falciparum* lifecycle [87, 88] with a majority of discrepancies attributed to a delay between transcript production and protein accumulation. Analysis of the *P. falciparum* proteome [94] and a comprehensive and integrated analyses of the genome, transcriptome and proteome of *P. berghei* and *P. chabaudi chabaudi*, which represents the state-of-the-art of functional genomics applied to the lifecycles of *Plasmodia* [95, 96], indicated that over half of the proteins in these parasites were detected solely in one stage of the lifecycle. This implies a considerable degree of specialization at the molecular level to support the demanding developmental program and suggested a highly coordinated expression of *Plasmodium* genes involved in common biological processes.

No single protein has physiological meaning alone and part of its mechanistic importance depends on its interactions with other proteins and cellular components. In-depth understanding of the protein-protein interaction network (defined as interactome) can provide insights into the function of proteins, regulatory mechanisms and functional relationships of these (Figure 1). An extensive protein-interaction study of *P. falciparum* (nearing global-type analysis) combining interaction information with co-expression data and GO annotations indicated unique interactions, identified groups of interacting proteins implicated in various biological processes, and predicted novel functions to previously uncharacterized proteins [97]. Interestingly, comparison of the *Plasmodium* interactome with that from other organisms indicated a marked divergence with very little conservation with other protein complexes [98].

The abovementioned X-ome datasets are therefore now available for *in silico* data mining approaches. As such, data mining of the transcriptome using an extensive sequence similarity search identified 92 putative proteases in the *P. falciparum* genome, 88 of which are actively transcribed [99]. This strategy has also been applied to the kinase family [100]. The transcriptome datasets are nowadays extensively utilized in PlasmoDB, MPMP and other specialized sites [101]. PlasmoDB additionally has links to proteome data and also allows access to the interactome data. Proteome experiments based on 2D-gel electrophoresis can furthermore be aided by Plasmo2D software to allow the identification of proteins in such platforms [102].

<Figure 1>

An important question in mining different global X-omic datasets is how they can be compared. Draghici et al. [103] made clear that the inconsistencies between the various microarray platforms (*in situ* synthesized short oligos, longer oligos, spotted oligos or cDNAs) are so high that it is almost impossible for the moment to compare results from different platforms. How can a transcriptomic profile be compared to a proteomic profile, given the errors, the linearity of signals and the magnitude of variations of each method, and the biological stability and turnover of RNA transcripts and of proteins? Is the enzyme profile correlated with the metabolite profile [104]? A pragmatic solution is to avoid the multiplicity of methods and by agreeing on some standards [103, 105-107]. Although "global" invariant references might not exist, it is nevertheless worth trying to find in large gene expression matrices the most invariant (or less variant) genes [106]. This quest is of general concern and is currently one of the challenges proposed by the European Conferences on Machine Learning and the European Conferences on Principles and Practice of Knowledge Discovery in Databases [106, 109]. In the absence of absolute references and standards, outputs from malaria X-omic experiments should be analyzed and compared with caution, particularly when obtained with platforms that are distinct from those used to feed the public data repositories. Consequently, in addition to referential public repositories for functional genomics, versatile software for local analyzes are strongly needed.

Various tools aim to provide biological interpretation of gene clusters but these mostly specialize in only one or two types of analyses. FatiGO [110], GeneLynx [111] and Gostat [112] are powerful tools for Gene Ontology mining; GoMiner [113], MAPPFinder [114] and DAVID [115] use GO and metabolic pathway interpretation whereas GeneXPress [116] and MiCoViTo [117] use metabolic pathways and incorporate transcription regulation visualization. Improvements on these include a web interface called MADIBA (MicroArray Data Interface for Biological Annotation, [118]) that has been initially designed for malaria transcriptomics (Dr. C. Claudel-Renard, personal communication). This interface links a relational database of various data sets to a series of analysis tools designed to facilitate investigations of possible reasons for co-expression of clusters of genes (e.g. from gene expression data) and to deduce possible underlying biological mechanisms. Clusters of co-expressed genes are automatically subjected to five different analytical modules including 1) search for over-represented GO terms in clusters, 2) visualization of related metabolic graphs with KEGG representations, 3) chromosomal localization, 4) search for motifs in the upstream sequences of the genes and finally 5) *Plasmodium*-specific genes without human homologues. MADIBA analysis of the transcriptome dataset from Le Roch et al. [85] resulted in an improved annotation of the *Plasmodium* genes (41% vs. 37%) and characterization of 6 additional clusters with GO annotations, of which one exclusively contained glycolysis in its entirety (except for fructose-bisphosphate aldolase) and another identifying gene as potential as drug targets due to their *Plasmodium*-specific characteristics. Therefore, MADIBA allows versatile analyses of a vast variety of transcriptomic profiles. These analyses can highlight potential drug targets by providing functionality to co-clustered expressed genes in a guilt-by-association manner, including those of un-annotated proteins, by predicting co-regulated expression *via* chromosomal localization as well as the identification of motifs for *cis*-regulatory elements and lastly by identifying unique *Plasmodium*-specific genes involved in specific biological mechanisms.

With the advent of integrative investigations of datasets from the transcriptome, proteome, interactome etc. new analysis tools are being developed including a Partial Least Squares (Projection to Latent Structures–PLS) method which has been used to integrate yeast transcriptome and metabolome data [119]. Linear modeling was used to investigate the changes in the transcriptome due to environmental perturbations and, assuming that the metabolome is a function of the transcriptome, the metabolic variables were modeled with PLS. A genome-wide investigation of protein function was recently performed by computationally modeling the *P. falciparum* interactome [120] to elucidate local and global functional relationships between gene products. This novel approach entailed an integration of *in* *silico* and experimental functional genomics data within a Bayesian framework to create the network of pairwise functional linkages. This resulted in predicting functionality based on associations between characterized and uncharacterized proteins for 95% of the currently annotated hypothetical proteins in the *P. falciparum* proteome. Only 107 hypothetical proteins show interaction with other hypothetical proteins potentially representing new pathways or previously uncharacterized components of known pathways.

Less than 500 of the known pharmaceutical drug targets act in the mechanism of currently known drugs. This brief overview shows that the integration and mining of global functional genomic experiments are in the frontline in drug discovery processes [121, 122]. Malaria X-omics data provide comprehensive information to, amongst others, understand the mode-of-action (MOA) of inhibitory compounds, allow optimization of drug action, validate drug targets (chemical validation strategies), identify families of genes/gene products that are more amenable as drug targets ('druggable genes'), annotate the function of hypothetical proteins by 'guilt-by-association' and point out specialized gene expression regulation systems (See Figure 1) [123].

In the case of drug treatments, the sought effects on the metabolism of targeted tissues or organisms include up- or downregulation of the protein target(s), the upregulation of detoxification pathways (cytotoxic responses) and the upregulation of alternative or compensatory pathways of the affected organism that can be reflected in changes in the transcriptome/proteome of the organism. The characterization of the transcriptional response induced by drug challenge has been applied with success in the antibacterial field, creating reference compendia of expression profiles after drug challenge that provide insight into a drugs' MOA [123-125]. Transcriptional profiling of drug challenged malaria parasites has been limited to only a few studies to date including Serial analysis of gene expression (SAGE) of chloroquine treated parasites [126], a high-density short-oligonucleotide array study on parasites treated with phosphatidylcholine biosynthesis inhibitors [88] and custom arrays originating from suppression, subtractive hybridisation (SSH) libraries on parasites treated with polyamine biosynthesis inhibitors [127]. Drug-specific transcriptional responses were seen in the chloroquine and polyamine inhibition studies indicating the presence of a feedback signaling mechanism. Le Roch et al. [88] compared transcriptional responses with proteome analyses and showed that more pronounced changes were induced at the protein level after drug challenge. This is also true for antifolate inhibited parasites where a marked increase were seen in folate biosynthesis protein levels upon treatment with inhibitors against DHRF-TS [128]. Global-level proteome response analysis of the combination of artemether and lumefantrine also revealed drug-specific changes in the proteome [129]. Subproteomic investigations have additionally become particularly important to determine the molecular binding partner/target protein of an inhibitory compound and/or to describe the mode-of-action of such compounds. This has been applied to ferriprotoporphyrin IX were identified [130], kinase inhibitors [131] and the quinoline family of compounds [132].

Any of the abovementioned strategies are potential starting points to the discovery of unsuspected drug targets in *P. falciparum*, whether it is a new/additional functionality that is ascribed to a known protein or the characterization of novel function of a previously 'unknown' protein. As the hypothetical proteins represent more than half (~60%) of the malarial proteome (see above), these are some of the most attractive areas to the drug target discovery effort. The basic rationale behind using expression profiles to assign functionality to genes is based on the principle of guilt-by-association [133, 134] in which genes coding for proteins with similar functionality often exhibit the same expression profiles and protein-interacting partners. Coincidentally, the expression profiles of genes specific to a given organelle also display similar expression patterns. This principle has been applied with unsupervised robust k-means clustering [85]. Improvements on this clustering approach were proposed using a semi-supervised clustering method called ontology-based pattern identification (OPI) [135]. OPI uses previous gene annotation data to generate clusters with greater specificity and confidence whose members then additionally share the same expression profiles. However, Llinás and del Portillo [136] warns against using only classical guilt-by-association methods, showing that many genes that are functionally unrelated show similar expression profiles during the asexual development of *P. falciparum*.

### Malaria protein structures and virtual ligand screening on candidate targets

Vital malaria proteins may have no counterpart in humans or sequence dissimilarity with their human homologues that may be sufficient to become a therapeutical target without disturbing essential function in the human host. The literature on potential protein targets for anti-malaria treatments is already crowded, and will hopefully be enriched and better documented in the future. Once a target has been identified, an important decision is whether or not to enter into costly drug- or vaccine-development programs. Determination or prediction of the target three-dimensional structure and *in silico* experiments can be achieved to connect these biological targets with the pharmacological space of small compounds [137], and assist this decision and provide initial clues on possible therapeutical strategies.

#### Malaria protein structures.

The rationalized identification of new inhibitors depends on possession of structural information. As for any other organism, the primary problem is obtaining high and pure protein yields for crystallization trials. Recombinant expression of malarial proteins in *E. coli* is notoriously difficult, however. A number of problems are typically encountered. The A+T richness results in substantially different codon usage compared to *E. coli*. *Plasmodium* genes are also typically much longer than their homologues in other organisms, as are the resulting proteins. Increased protein size is due mostly to long protein inserts with generally little homology to cognate enzymes. These inserts tend to be disordered and of low complexity, resulting in proteins that are not amenable to expression and crystallization. Further problems include sporadic mutations of low complexity sequences introduced by *E. coli*, and cryptic prokaryotic translation start sites within malarial genes. Some level of protein expression may be obtained by fine control of expression conditions, often a change of strain or of complete expression system, addition of rare codon tRNAs, and more and more often by production of synthetic genes coding for identical protein sequences but with a codon usage optimized for bacteria [138-142]. Mehlin et al. [143] recently attempted a wholesale expression of 1000 malarial genes and obtained soluble expression for only 63 genes. High predicted disorder, molecular weight, pI and lack of homology to *E. coli* proteins were all negatively correlated with soluble expression.

The difficulty of expressing malarial proteins is reflected by the paucity of structures in the Protein Data Bank [144]. At the time of writing there are only 64 non-redundant *Plasmodium* protein structures in the PDB (Table 2). In contrast, querying the PDB for human entries (excluding > 90% sequence identity) reveals more than 1700 structures. The advent of structural genomics programs (the Structural Genomics Consortium, [145]; the Structural Genomics of Pathogenic Protozoa, [146]) has increased the throughput of new malarial structures. Since 2003, 56 depositions of *Plasmodium* structures have been made. Whether this trend will continue beyond the "low hanging fruit" remains to be seen.

<Table 2>

In lieu of crystal structures for malarial proteins many groups have resorted to homology modeling. This approach depends critically on the alignment with template structures. Unfortunately the biased nucleotide and amino acid composition (see above and [58]) and *Plasmodium*-specific inserts make it difficult to correctly identify core-conserved regions. The presence of inserts often confuses multiple and structural-alignment programs. A number of techniques have been used to circumvent this problem (see Figure 2). From a first pass alignment, approximate insert positions can be determined. Sequences can then be split according to long inserts and re-aligned. Inserts can vary considerably across different *Plasmodium* species ([147] and C. Claudel-Renard, personnal communication). Therefore, while adjusting an alignment for modeling, it is useful to refer to phylogenetically diverse multiple alignments including as many *Plasmodium* sequences as possible (see above, [148]). As an adjunct to alignment, independent motif identification (*e.g.* the MEME system; [149, 150]) can be used to fix mistakes that alignment programs frequently make when aligning long *Plasmodium* proteins with homologues [148, 151]. Further improvements can be made by using hydrophobic cluster analysis [61] and secondary structure predictions to align homologous regions within inserts. Once an alignment has been decided on, based on visual assessment, a series of models can be built. Because of the high degree of uncertainty that often accompanies alignments used for modeling malarial proteins, it is usually not feasible to rectify all structural anomalies. But by performing standard quality checks on a large sample of models and summarizing the results, it is possible to identify parts of the alignment causing most problems. Refined alignments might benefit from species-specific matrices that take into account the differences of amino acid distribution between the aligned proteins [60, 152].

<Figure 2>

Despite the difficulties with homology modeling of malarial proteins there have been some notable successes. Malarial DHFR forms part of a bifunctional protein that also carries thymidylate synthase. A number of existing drugs such as cycloguanil and pyrimethamine target the DHFR domain, and have been used effectively in the past. However drug resistance has evolved that reduces the usefulness of this important class of drugs. Hence malarial DHFR has been a popular target for homology modeling efforts [153-158]. Toyoda et al. [153] were able to identify new inhibitors in the micromolar range. McKie et al. [154] and Lemcke et al. [155] could rationalise the pyrimethamine resistance caused by the S108N mutation. One of these models was further used to identify new inhibitors acting in the nano- and micromolar ranges [154]. Delfino et al. [158] in turn used their model to investigate a large number of antifolate resistant mutants. Rastelli et al. [156] further explained the cycloguanil resistance/pyrimethamine sensitivity conferred by A16V+S108T, as well as the ability for WR99210 to inhibit both pyrimethamine and cycloguanil resistant mutants. A number of new inhibitors were also successfully designed. The high accuracy of the alignment used for modelling meant that predicted dockings were subsequently confirmed with the crystal structure of the complete bifunctional enzyme [159]. Considerable work has also gone into modeling malarial proteases essential to the parasite's intra-erythrocytic life stage. A number of these models have been used to identify new inhibitors [160-163], although the increasing number of crystal structures for these proteases is likely to gradually replace the need for homology models (Table 2).

Apart from the PDB, there is currently no resource that includes all *Plasmodium* protein structures. Both publicly available versions of PlasmoDB (4.4 and 5) are still incomplete in this regard. PlasmoDB 4.4 only lists *P. falciparum* structures, and there is incomplete overlap between versions 4.4 and 5. PlasmoDB 4.4 includes a section for >440 modeled structures based on a wholesale modeling attempt of >5000 open reading frames documented at [164]. However, at the time of writing the actual structures were not available. Furthermore, PlasmoDB 5 also lists non-*Plasmodium* structures, and therefore requires some curating. We argue that a dedicated resource for deposition of structures from the *Plasmodium* structural community would be useful. The resource should also include model structures subjected to the same rigor as experimental structures in the PDB, including quality criteria and scoring. Thus "experimental" details for modeling (alignments, software methods etc) should be included. This resource should also allow easy comparison with corresponding solved structures enabling evaluation of the communities' ability to model this difficult organism.

Protein structures provide invaluable information for drug or vaccine discovery. First, compliance of the structure with known properties of recorded pharmaceutical targets (termed "druggable" genes) [165-167] or prediction of the occurrence of immunogenic epitopes (termed "immunizable" genes) [168-169], can be investigated. Second, having a protein structure in hand is a very early, but necessary step, for the *in silico* prediction of the families of ligands that can interfere with protein function.

### Virtual ligand screening.

Advances in combinatorial chemistry have broken limits in organic synthetic chemistry and accelerated the production throughput. Thus, millions of chemical compounds are currently available in private and academic laboratories and recorded in 2D or 3D electronic databases. It is often technically impossible and very expensive to screen such a high number of compounds using wet high throughput screening techniques. An alternative is high throughput virtual screening by molecular docking, a technique which can screen millions of compounds rapidly and cost effectively. Molecular docking is computer based method which predicts the ligand conformations inside the active site of the target as well as an estimate of the binding affinity between protein and ligand. It also gives insight about the interactions between protein and ligand and allows to generate mode-of-action hypotheses. Screening each compound, depending on its structural complexity, requires from a few seconds to hours of computation time on a standard PC workstation depending on the chosen docking algorithm. Consequently, screening all compounds in a single database would require years. However, the problem is embarrassingly parallel and the computation time can be reduced very significantly by distributing data to process over a grid gathering thousands of computers [170, 171].

Recently, virtual screening projects on grids have emerged with the purpose of reducing cost and time. They focused on the development of an *in silico* docking pipeline on grids of clusters [172] but also on the optimization of molecular modelling [173]. Pharmaceutical laboratories have also become interested by the grid concept; Novartis deployed the first automated modelling and docking pipeline on an internal grid [174]. Other projects focused on virtual screening deployment on a pervasive grid, or desktop grid, to analyse specific targets [175].

In mid-2005, the WISDOM (World-wide In Silico Docking On Malaria) initiative successfully deployed large scale *in silico* docking on the European public EGEE grid infrastructure [176]. The biological targets were plasmepsins, aspartic proteases of *Plasmodium* responsible for the initial cleavage of human haemoglobin [177]. There are ten different plasmepsins coded by ten different genes in *P. falciparum* (Plm I, II, IV, V, VI, VII, VIII, IX, X and HAP) [178]. High levels of sequence homology are observed between different plasmepsins (65-70%). Simultaneously they share only 35% sequence homology with their nearest human aspartic protease, Cathepsin D4 [179]. This and the presence of accurate X crystallographic data made plasmepsin an ideal target for rational drug design against malaria.

The main goal of the WISDOM project has been to make use of the EGEE grid infrastructure to set up an *in silico* experimentation environment. Having the necessary computing power at hands, scientists from a virtual organization can design new large scale test systems for generating new hypotheses. The benefit is that a large number of targets can be combined with a very large number of potential hit molecules, using different docking algorithms and allowing to chose different parameter settings. As discussed above for the X-omic experiments in post-genomic platforms, *in silico* methods pose the same important question in order to mine the data, *i.e.* how they can be compared. The careful input data preparation is a crucial step in the process, which has to be performed by experts and be made available for the whole scientific community. By sharing the results of virtual screenings in a common knowledge space, different experts coming from different fields can jointly derive a rational for selecting appropriate combinations of targets, ligands and virtual screening methods.

The WISDOM large scale *in silico* docking deployment saw over 46 million docked ligand-protein solutions, resulting from 2 docking tools, 5 targets, 1 million compounds and 4 parameter settings, the equivalent of 80 years on a single PC in about 6 weeks. Up to 1700 computers were simultaneously used in 15 countries around the world. Post-processing of the huge amount of data generated was a very demanding task as millions of docking scores had to be compared. At the end of the large scale docking deployment, the best compounds were selected based on the docking score, the binding mode of the compound inside the binding pocket and the interactions of the compounds to key residues of the protein.

Several promising scaffolds have been identified among the 100 compounds selected for post processing. Among the most significant ones are urea-, thiourea-, and guanidino analogues, as these scaffolds are most repeatedly identified in the top 1000 compounds (see Figure 3). Validating this approach, some of the compounds identified were similar to already known plasmepsin inhibitors, like urea analogues from the Walter Reed chemical database, which were previously established as micro molar inhibitors for plasmepsins [180]. This indicates that the overall approach is sensible and large scale docking on computational grids has real potential to identify new inhibitors. In addition, guanidino analogues appeared very promising and most likely to become a novel class of plasmepsin inhibitors.

<Figure 3>

The developed and established protocols can be used for coming scenarios. Several teams have expressed interest to propose targets for a second computing challenge called WISDOM II that should be carried out in late 2006. While docking methods have been significantly improved in the last years, docking results need to be post-processed with more accurate modeling tools before biological tests are undertaken. The major challenges for docking methods still are pose prediction and scoring. Molecular dynamics (MD) has great potential at this stage: firstly, it enables a flexible treatment of the ligand/target complexes at room temperature for a given simulation time, and therefore is able to refine ligand orientations by finding more stable complexes; secondly, it partially solves conformation and orientation search deficiencies which might arise from docking; thirdly, it allows the re-ranking of molecules based on more accurate scoring functions. Efforts are now devoted to deploy on the grid both docking and Molecular Dynamics calculations to further accelerate *in silico* virtual screening before in vitro testing.

## Conclusions: toward a chemogenomic knowledge space

In this paper, we examined five aspects of the *in silico* storage, organization and mining of data from malaria genomics and post-genomics, in the context of the prediction and characterization of targets and drugs: 1) the comparison of protein sequences including compositionally atypical malaria sequences, 2) the high throughput reconstruction of molecular phylogeny, 3) the representation of biological processes particularly metabolic pathways, 4) the versatile methods to integrate genomic data, biological representations and functional profiling obtained from X-omic experiments after drug treatments, 5) the determination and prediction of protein structures and their virtual docking with drug candidate structures. We focused our attention on the data management that should be at the genomic scale, including multiple species, and should therefore be as reliable as possible. We were concerned that a "biological space" should be formatted so as to represent our knowledge and to connect genomic data and post-genomic functional profiles in the most versatile way, allowing the mining of information with diverse methods (Figure 4a-e). This "biological space" should be linked to a "chemical space" that contains all small molecules stored in chemolibraries (millions of compounds) including known drugs (Figure 4f-h). Thus, we argue that progresses toward a chemogenomic knowledge space can benefit on the referential public repositories and particularly UniProt and PlasmoDB, on recent theoretical advances in genomics and post-genomics data management and mining and on the power of computer grids (Figure 4).

<Figure 4>

Genomic data (Figure 4a), *i.e.* protein sequences, can be organized based on sequence similarity (Figure 4b). This projection of protein sequences should allow the high throughput reconstruction of molecular phylogenies both at the intraspecific (connecting paralogs and alleles) and interspecific (connecting homologues among which orthologs) levels, following statistically accurate methods (Figure 4b). This task is difficult because of the compositional bias and high insert content of malaria sequences. Statistically valid protein sequence comparisons are now available to allow genome scale alignment and high throughput phylogeny reconstruction (named TULIP), including malaria atypical sequences, and providing quality scores on which one can rely after automatic genome scale treatments. Benefits from the TULIP method include an easy upgrade and update of the obtained protein spatial projection. Another substantial side of a biological space contains representations of our knowledge of biological processes, using stable ontologies, and dynamic graph representation (Figure 4c). Here, efforts are still needed to improve existing representations of the metabolism and numerous projects are under progress, most importantly PlasmoCyc and PMPM. The PlasmoCyc existing metabolic graphs have the advantage of being more easily updated and usable for *in silico* mining methods, as long as the output is examined by biological experts. In a brief overview of malaria X-omic profiling, we introduced some tools allowing a linkage between genomic data, biological process representations and global functional profiles (Figure 4d). There is still a large diversity of data treatment and mining strategies, reflecting the diversity of technologies and mining methods. This step is one of the doors to connect the biological space with the chemical space, following the "reverse chemical genetic" way, *i.e.* "from known drugs to biological response" (toxicity, mode of action). Basic analytical tools like MADIBA and sophisticated mining approaches will be needed to understand and compare the biological responses to anti-malarial drugs. The other door to connect the biological space and the chemical space follows the "direct chemical genetic" way, *i.e.* "from known biological target to drug candidate". We argue that in addition to crystal structures of malaria proteins, the automated structural annotation of the malaria proteome should be initiated (Figure 4e). Based on protein structure information, virtual docking campaigns such as the WISDOM challenges can be achieved using the power of computer grids.

This paper did not review the "chemical" side for chemogenomics space. By numerous aspects the *in silico* organization of the small molecules stored in chemolibraries (Figure 4f) was not achieved the way biological information was. Clustering of small molecular structures based on properties is highly debated (Figure 4h). Collections of small molecules are generally designed to obey the pragmatic Lipinski's "rule of five" [181] making them likely candidates for drug discovery. Numerous studies are under progress in order to identify which small molecule descriptors can comply with chemogenomic approaches (*e.g.* the Accamba project for the analysis of chemolibraries and the building of bioactivity models; [182]). A chemical ontology (CO) has been recently introduced [183], but it has not yet been used and validated by the scientific community the way the GO was. A database for Chemical Entities of Biological Interest (ChEBI) has also been launched [184]. The modeling of the three-dimensional structures of small molecules (Figure 4g) can be predicted by numerous public or commercial methods which should be examined with attention (see the WISDOM challenges). PubChem, a repository for molecules acting on biological targets was recently launched [183, 185] and the UniProt protein knowledge base was recently upgraded to report toxic doses of small molecules on proteins [186, 187], however these initiatives are just starting points. Access to an ocean of small molecular structures and to a deluge of biological sequences raises an enthusiastic challenge: “The goal for the coming decades will be to explore the overlap between chemistry space and protein space” [188]. Is this prediction exuberant? From the methodological survey summarized in this paper, we believe that this next milestone for malaria research is not out of reach.

Beyond virtual screening, the grid technology provides the collaborative IT environment to enable the coupling between molecular biology research and goal-oriented field work [189]. It proposes a new paradigm for the collection and analysis of distributed information where data are no more to be centralized in one single repository. On a grid, data can be stored anywhere and still be transparently accessed by any authorized user. The computing resources of a grid are also shared and can be mobilized on demand so as to enable very large scale genomics comparative analysis and virtual screening. A longer term perspective is therefore to enhance the ability to share diverse, complex and distributed information on a given disease for collaborative exploration and mutual benefit. The concept of a knowledge space is to organize the information so that it can be reached in a few clicks. This concept is already successfully used internally by pharmaceutical laboratories to store knowledge [175]. The grid permits the building of a distributed knowledge space so that each participant is able to keep the information he owns on his/her local computer. A set of grid services would particularly take advantage of the developments in the area of semantic text analysis for extraction of information in biology and genome research. (Figure 4, lower part). The literature on malaria biology, physiopathology and medicinal chemistry is, at least in significant parts, stored as unstructured texts that make an invaluable source of knowledge which access depends on advances in terminology analysis and term extraction methods. A first attempt at using terminology analysis for the "harvesting" of relevant concepts in a defined disease area has been undertaken in another biomedical grid project, the recently started EU Integrated Project @neuRIST [190]. In this project, terminology analysis lead to the refinement of a disease-specific text corpus and provided a shortlist of relevant terms. Moreover, not only the genes associated with the defined disease area could be identified by automated methods, but also single nulceotide polymorphisms (SNPs) published for these disease-associated genes, could be automatically identified (Dr. Laura Furlong, IMIM, Barcelona, personal communication). An important focus of future activities in this project is the construction and validation of fine-granular disease-specific ontologies. This concept can easily be adopted for a knowledge base on any disease, including malaria. Existing databases can be complemented by this automatically generated semantic layer, which subsequently would also be helpful for data mediation. Moreover, a structured knowledge space would produce grid services for indexing of distributed data resources and thus improve navigation through knowledge and retrieval of relevant information.

Finally, present and future *in silico* information must be supported and validated by data gathered *in vitro* and *in vivo*. The challenges in this domain remain severe. Difficulties of cloning and expressing parasite proteins in heterologous systems, the validation of druggable targets using siRNA etc, and the experimental assignment of functionality to the many hypothetical proteins, will occupy scientists in parallel for some time to come. Data generated from *in silico* analysis leads to a need for further laboratory work, such as the *in vitro* testing of ligands identified as potential drug leads through the WISDOM project. A strong linkage between scientists undertaking *in vitro* research and *in silico* researchers is therefore essential to support an iterative approach to knowledge generation and analysis, in the context of malaria chemogenomics.

## Authors' contributions

All authors contributed to the analysis of the current status on the *in silico* storage and organization of malaria genomics, post-genomics data and antimalaria chemical information, as made available in the literature and public websites. All authors contributed to the writing of the current review manuscript.

## Acknowledgements

We thank Dr. Jane Morris for discussions and in-depth reading of this manuscript.

## Figures

#### Figure 1 - Malaria functional genomics (X-omics) strategies in the context of target and drug characterization

Selected questions that could be addressed by the application of functional genomics are listed, including those specific to the transcriptome, proteome or interactome (X-omes). Highlighted is the particular focus on the application of this type of strategies to drug(s) and target(s) characterization.

#### Figure 2 - Current pipeline for the homology-based modeling of malaria protein 3D-structures

This scheme emphasizes on the currently available methods to overcome amino acid bias, low sequence identity, protein inserts etc. Future upgrades include the refinement of each of these methods, for instance implementing asymmetric substitution matrices discussed in the text, that take into account the different amino acid distributions of malarial and non-malarial proteins for pairwise alignments.

#### Figure 3 - *In silico* screening for protein ligands based on structural docking

A urea analogue inhibiting malaria plasmepsins was identified with good score from the first WISDOM (World-wide In Silico Docking On Malaria) campaign. The WISDOM initiative successfully deployed large scale *in silico* docking on the European public EGEE grid infrastructure. The ligand shown here docks inside the binding pocket of plasmepsin, and interacts with key protein residues. The developed and established protocols can be used for new targets, and particularly a second computing challenge, WISDOM II.

#### Figure 4 - Malaria chemogenomics: organization and treatment of genomic, post-genomic and chemical information for the prediction and characterization of target and drug candidates

Genomic data from *Plasmodium* and other species (a), *i.e.* protein sequences, should be organized based on sequence similarity (b). This projection should allow the high throughput reconstruction of molecular phylogenies both at the intraspecific (connecting paralogs and alleles) and interspecific (connecting homologues among which orthologs) levels following statistically accurate methods *e.g.* the TULIP method. Another substantial side of the biological space is designed by representing our knowledge of the biological processes, using stable ontologies *e.g.* the GO, and dynamic graph representation, *e.g.* PlasmoCyc (c). Versatile tools should allow the integration of genomic data, biological process representations and global functional profiles obtained with diverse X-omic approaches (4). These tools should comply with the large diversity of technologies and mining methods. The collection of information on the biological response to drugs is one of the doors to connect the biological space with the chemical space, following the "reverse chemical genetic" way, *i.e.* "from known drugs to biological response" (toxicity, mode of action). The other door to connect the chemical space and the biological space follows the "direct chemical genetic" way, *i.e.* "from known biological target to drug candidates". In addition to malaria protein structures obtained from crystals, the automated structural annotation of the malaria proteome should be initiated with quality scores (e). Based on protein structure information, virtual docking campaigns such as the WISDOM challenges can be achieved using the power of computer grids. The *in silico* organization of the small molecules stored in chemolibraries (f) follows similar principles, in particular the determination of three-dimensional structures of small molecules (g) and a clustering of small molecular structures based on drug properties and descriptors (h). Sharing and mining of chemogenomic information, completed with knowledge harvested in unstructured scientific literature, would benefit of the advances in knowledge space design and deployment on knowledge grids.

# Tables

### Table 1 - Comparison of *Plasmodium falciparum*, *Saccharomyces cerevisiae*, *Arabidopsis thaliana* and *Homo sapiens* genomic statistics

Presented data compile information from [22] for *Plasmodium falciparum*, [190] for yeast (completed with statistics made available *via* the Comprehensive Yeast Genome Database website, [191]), the Arabidopsis genome initiative [192] for Arabidopsis, and the International Human Genome Sequencing Consortium [193] and [194] for Human (completed with statistics made available *via* Ensembl, [195]). These statistics at the complete genome release date have been continuously updated since then.

### Table 2 - Non-redundant malarial structures in the Protein Data Bank (PDB).

All entries were compiled from the PDB, and were cross-checked against PlasmoDB ([196], versions 4 and 5), the Structural Genomics Consortium, [145] and the Structural Genomics of Pathogenic Protozoa, [146]). *P. b.*, *Plasmodium berghei*; *P. c.*, *Plasmodium cynomolgi*; *P.f.*, *Plasmodium falciparum*, *P. k.*, *Plasmodium knowlesii*; *P. m.*, *Plasmodium malariae*, *P. v.*, *Plasmodium vivax*

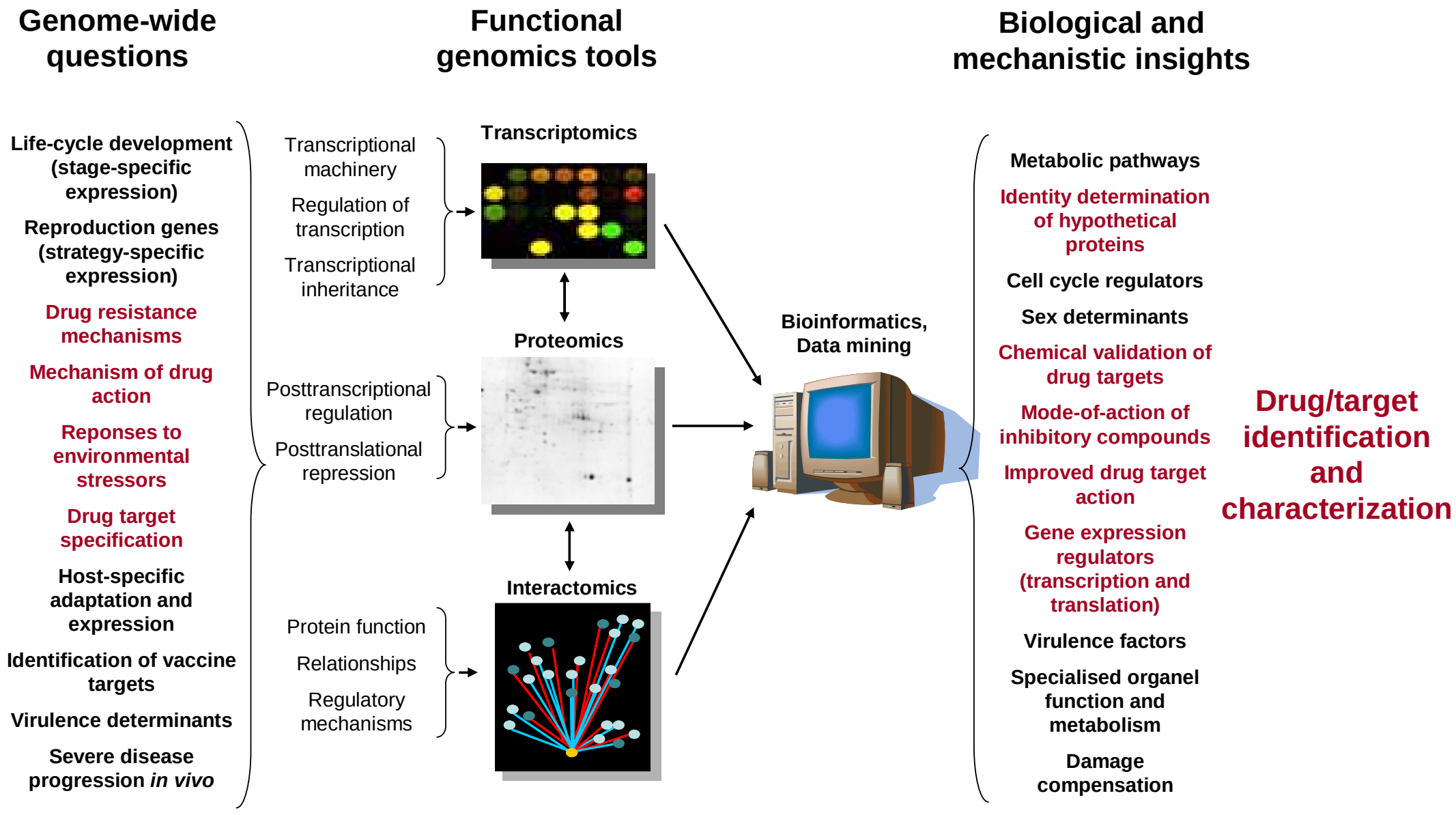
Genome-wide questions
Functional genomics tools
Biological and mechanistic insights
Life-cycle development (stage-specific expression)
Reproduction genes (strategy-specific expression)
Drug resistance mechanisms
Mechanism of drug action
Reponses to environmental stressors
Drug target specification
Host-specific adaptation and expression
Identification of vaccine targets
Virulence determinants
Severe disease progression in vivo
Transcriptional machinery
Regulation of transcription
Transcriptional inheritance
Transcriptomics
Proteomics
Posttranscriptional regulation
Posttranslational repression
Interactomics
Protein function
Relationships
Regulatory mechanisms
Bioinformatics, Data mining
Metabolic pathways
Identity determination of hypothetical proteins
Cell cycle regulators
Sex determinants
Chemical validation of drug targets
Mode-of-action of inhibitory compounds
Improved drug target action
Gene expression regulators (transcription and translation)
Virulence factors
Specialised organel function and metabolism
Damage compensation
Drug/target identification and characterization

Figure 1

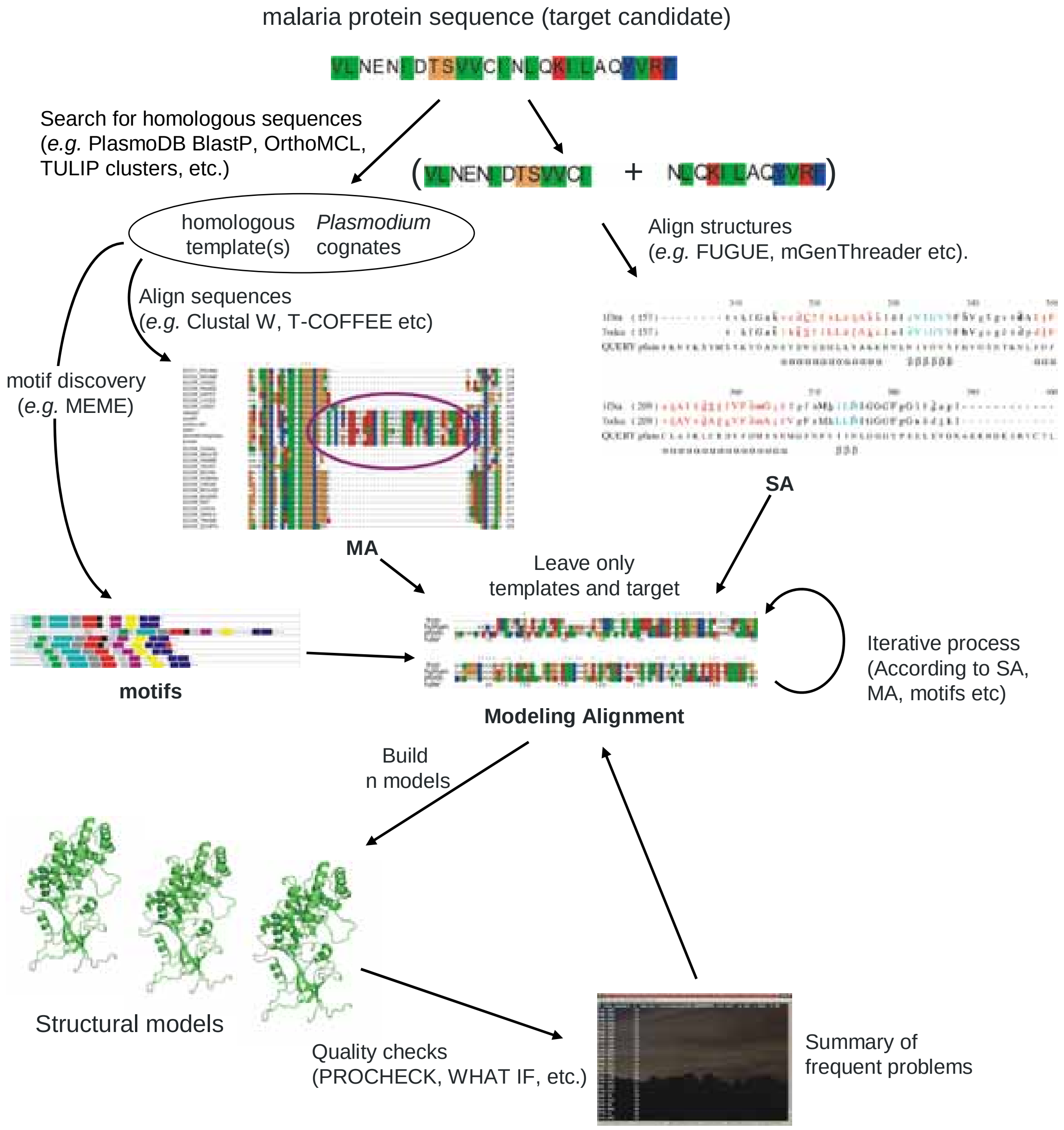
malaria protein sequence (target candidate)
Search for homologous sequences
(e.g. PlasmoDB BlastP, OrthoMCL, TULIP clusters, etc.)
homologous template(s)
Plasmodium cognates
Align structures
(e.g. FUGUE, mGenThreader etc).
Align sequences
(e.g. Clustal W, T-COFFEE etc)
motif discovery
(e.g. MEME)
SA
MA
Leave only templates and target
motifs
Modeling Alignment
Iterative process
(According to SA, MA, motifs etc)
Build
n models
Structural models
Quality checks
(PROCHECK, WHAT IF, etc.)
Summary of frequent problems

Figure 2

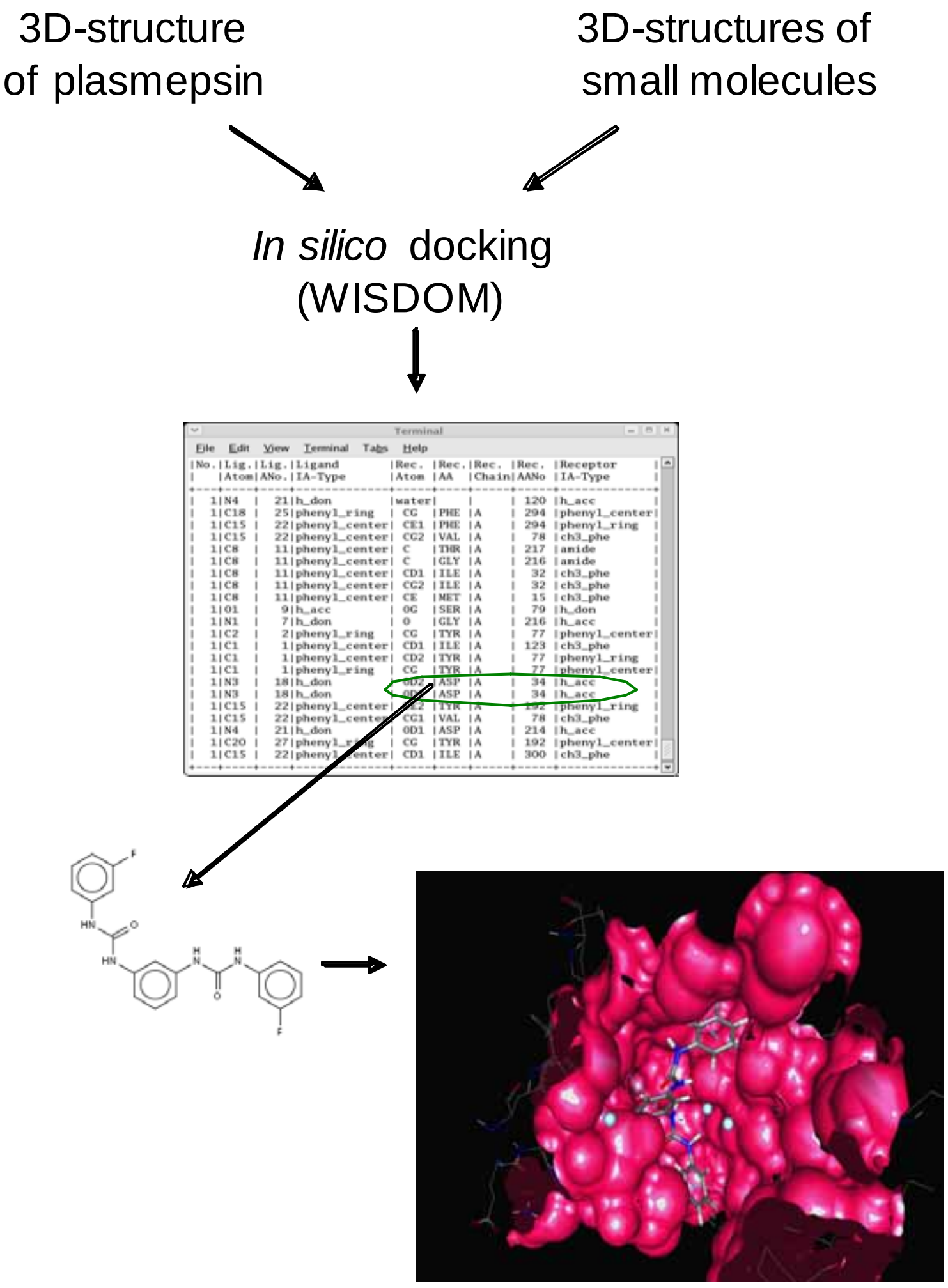
3D-structure of plasmepsin
3D-structures of small molecules
*In silico* docking (WISDOM)
Terminal
File Edit View Terminal Tabs Help
|No.|Lig.|Lig.|Ligand |Rec. |Rec.|Rec. |Rec. |Receptor |
| |Atom|ANo.|IA-Type |Atom |AA |Chain|AANo |IA-Type |
| 1|N4 | 21|h_don |water| | | 120 |h_acc |
| 1|C18 | 25|phenyl_ring | CG |PHE |A | 294 |phenyl_center|
| 1|C15 | 22|phenyl_center| CE1 |PHE |A | 294 |phenyl_ring |
| 1|C15 | 22|phenyl_center| CG2 |VAL |A | 78 |ch3_phe |
| 1|C8 | 11|phenyl_center| C |THR |A | 217 |amide |
| 1|C8 | 11|phenyl_center| C |GLY |A | 216 |amide |
| 1|C8 | 11|phenyl_center| CD1 |ILE |A | 32 |ch3_phe |
| 1|C8 | 11|phenyl_center| CG2 |ILE |A | 32 |ch3_phe |
| 1|C8 | 11|phenyl_center| CE |MET |A | 15 |ch3_phe |
| 1|O1 | 9|h_acc | OG |SER |A | 79 |h_don |
| 1|N1 | 7|h_don | O |GLY |A | 216 |h_acc |
| 1|C2 | 2|phenyl_ring | CG |TYR |A | 77 |phenyl_center|
| 1|C1 | 1|phenyl_center| CD1 |ILE |A | 123 |ch3_phe |
| 1|C1 | 1|phenyl_center| CD2 |TYR |A | 77 |phenyl_ring |
| 1|C1 | 1|phenyl_ring | CG |TYR |A | 77 |phenyl_center|
| 1|N3 | 18|h_don | OD2 |ASP |A | 34 |h_acc |
| 1|N3 | 18|h_don | |ASP |A | 34 |h_acc |
| 1|C15 | 22|phenyl_center| |TYR |A | 192 |phenyl_ring |
| 1|C15 | 22|phenyl_center| CG1 |VAL |A | 78 |ch3_phe |
| 1|N4 | 21|h_don | OD1 |ASP |A | 214 |h_acc |
| 1|C20 | 27|phenyl_ring | CG |TYR |A | 192 |phenyl_center|
| 1|C15 | 22|phenyl_center| CD1 |ILE |A | 300 |ch3_phe |

Figure 3

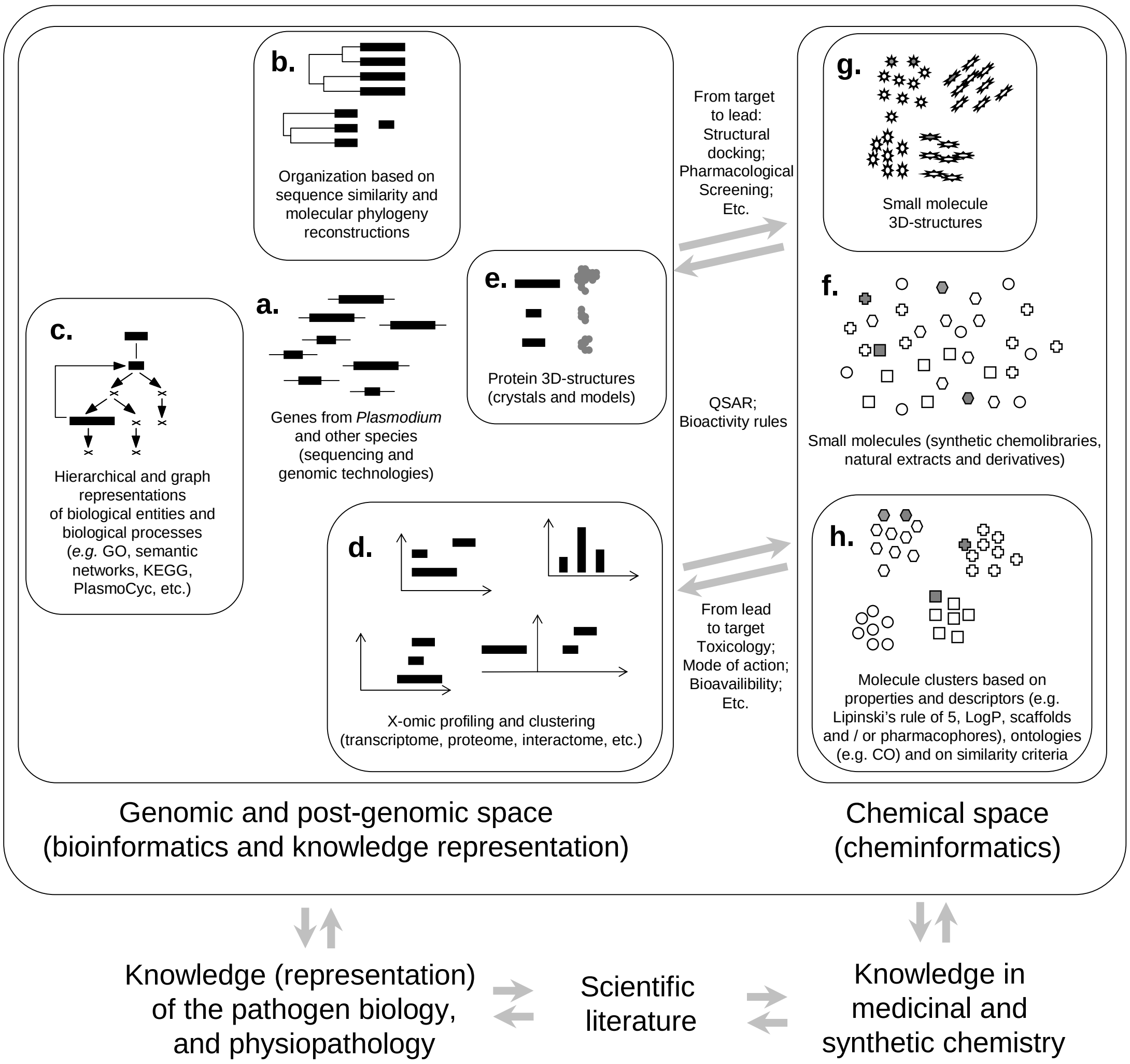
b.
Organization based on sequence similarity and molecular phylogeny reconstructions
a.
Genes from *Plasmodium* and other species (sequencing and genomic technologies)
e.
Protein 3D-structures (crystals and models)
c.
Hierarchical and graph representations of biological entities and biological processes (*e.g.* GO, semantic networks, KEGG, PlasmoCyc, etc.)
d.
X-omic profiling and clustering (transcriptome, proteome, interactome, etc.)
From target to lead: Structural docking; Pharmacological Screening; Etc.
QSAR; Bioactivity rules
From lead to target Toxicology; Mode of action; Bioavailibility; Etc.
g.
Small molecule 3D-structures
f.
Small molecules (synthetic chemolibraries, natural extracts and derivatives)
h.
Molecule clusters based on properties and descriptors (e.g. Lipinski's rule of 5, LogP, scaffolds and / or pharmacophores), ontologies (e.g. CO) and on similarity criteria
Genomic and post-genomic space (bioinformatics and knowledge representation)
Chemical space (cheminformatics)
Knowledge (representation) of the pathogen biology, and physiopathology
Scientific literature
Knowledge in medicinal and synthetic chemistry

Figure 4

| | *Plasmodium falciparum* | *Saccharomyces cerevisiae* | *Arabidopsis thaliana* | *Homo sapiens* |
|---|---|---|---|---|
| **Genome general statistics** | | | | |
| No of chromosomes | 14 | 16 | 5 | 22 + X/Y |
| Size (bp) | 22,853,764 | 12,495,682 | 115,409,949 | 3,272,187,692 |
| average (A+T) % | 80.6 | 61.7 | 65.1 | 59.0 |
| Estimated number of genes | 5,268 | 5,770 | 25,498 | 31,778 |
| Average gene length | 2,283 | 1,424 | 1,310 | 1,340 |
| % of coding genome | 53 | 66 | 29 | 9 |
| **Initial annotation based on sequence similarity (BLAST or *Smith-Waterman *E-values*)** | | | | |
| Proportion of predicted protein sequences: | | | | |
| - having a detectable similarity to sequences, in other organisms, of known function at the initial genome release date. | 34 % | 75 % | 69 % | 59 %* |
| - without any detectable similarity to sequences in other organisms at the initial genome release date, i.e. "no BLASTP match to known proteins" (estimates based on published data and local BLAST searches). | 61 % | < 8 % | < 20 % | 15 %* |
| - of totally unknown function (hypothetical proteins = with similarity to sequences of unknown function + without any detectable similarity to sequences in other organisms). | 66 % | 16 % | 31 % | 41 %* |
| **Average characteristics of open reading frames** | | | | |
| Exons: | | | | |
| No per gene | 2.39 | 1.05 | 5.18 | 12.1 |
| (A+T) % | 76.3 | 60 | 55 | 52 |
| average length | 949 | 1356 | 253 | 111 |
| Introns: | | | | |
| (A+T) % | 86.5 | 64 | 66 | 60 |
| Intergenic regions: | | | | |
| (A+T) % | 86.4 | 64 | 66 | 60 |

Table 1

| Description | PDB Ids | Ligands | Species |
|---|---|---|---|
| 1-Cys peroxiredoxin | 1XIY, 1XCC | | *P. f., P. y.* |
| 2-Cys peroxiredoxin | 2H66, 2FEG | | *P. v.* |
| 6-pyruvoyl tetrahydropterin synthase | 1Y13, 2AOS | Biopterin | *P. f., P. v.* |
| Acyl CoA binding protein | 1HBK | | *P. f.* |
| Adenylosuccinate synthetase | 1P98 | | *P. f.* |
| Adenylosuccinate lyase | 2HVG | | *P. v.* |
| Apical membrane antigen 1 | 1HN6, 1Z40, 1W81, 1W8K | | *P. f., P. v.* |
| beta-hydroxyacyl-ACP dehydratase | 1Z6B, 1ZHG | Cacodylate | *P. f.* |
| Cell division control protein 2 homolog, Pfpk5 | 1V0P, 1VOB, 1V0O, 1OB3 | Purvalanol B, Indirubin-5-Sulphonate | *P. f.* |
| ClpP protease | 2F6I | | *P. f.* |
| C-terminal merozoite surface protein | 1B9W | | *P. c.* |
| cyclophilin, peptidyl-prolyl cis-trans isomerase | 1QNH, 1QNG, 1Z81, 2B71, 2FU0 | Cyclosporin A | *P. f., P. y.* |
| DHFR-TS | 1J3J, 1J3I, 1J3K, 2BL9, 2BLB | Pyrimethamine, NADPH , dUMP, WR99210, NDP, CP6, MES, CP7 | *P. f., P. v.* |
| Dihydroorotate dehydrogenase | 1TV5 | orotic acid, n8e, fmn, a26 | *P. f.* |
| Dimethyladenosine transferase | 2H1R | | *P. f.* |
| dUTPase | 1VYQ | 2 ,3-deoxy-3-fluoro-5-o-trityluridine | *P. f.* |
| Dynein Light Chain 1 | 1YO3 | | *P. f.* |
| Enoyl-acyl-carrier protein reductase | 1NHG, 1NHW, 1NHD/ 1VRW, 1NNU | Triclosan, NAD+, TCC, TCT | *P. f.* |
| Falcipain-2 | 1YVB, 2GHU | | *P. f.* |
| Ferredoxin | 1IUE | | *P. f.* |
| Fe-superoxide dismutase | 2AWP, 2A03 | | *P. k., P. b.* |
| Fructose-bisphosphate aldolase | 1A5C | | *P. f.* |
| Glutamate dehydrogenase | 2BMA | | *P. f.* |
| Glutathione reductase | 1ONF | | *P. f.* |
| Glutathione S-transferase | 1PA3, 1Q4J, 1OKT, 2AAW | S-Hexyl-Gsh, P33, GTX, DTL | *P. f.* |
| Glyceraldehyde-3-phosphate dehydrogenase | 1YWG, 2B4R, 2B4T | NAD+, glycerol, AES | *P. f.* |
| Glycerol-3-phosphate dehydrogenase | 1YJ8 | | *P. f.* |
| Guanylate kinase | 1Z6G | 4-(2-hydroxyethyl)-1-piperazine ethanesulfonic acid | *P. f.* |
| Hypothetical protein | 1ZSO | | *P. f.* |
| Hypoxanthine phosphoribosyl-transferase | 1CJB | | *P. f.* |
| L-lactate dehydrogenase | 1LDG, 1CET, 1CEQ, 1T26, 1T2E, 1T25, 1T24, 1T2D, 1T2C, 1U4O, 1U5S, 1U5C, 1U5A, 1XIV, 2A94, 1OC4, 2A92, 2AA3 | NADH,Oxamate, Chloroquine, 4-Hydroxy-1,2,5-Thiadiazole-3-Carboxylic Acid, 3-Hydroxyisoxazole-4-Carboxylic Acid, NAD+, 4-Hydroxy-1,2,5-Oxadiazole-3-Carboxylic Acid, 2,6-dicarboxynaphthalene, naphthalene-2,6-disulfonic acid, 3,7-dihydroxynaphthalene-2-carboxylic acid, RB2, glycerol, Acetyl pyridine adenine dinucleotide, 1,4-dihydronicotinamide adenine dinucleotide, ADAPH | *P. f., P. b., P. v.* |
| MDR 2 | 2GHI | | *P. y.* |
| MSP1 | 1OB1, 2FLG, 1CEJ, 1N1I | | *P. f., P. k.* |
| MSP3 | 1PSM | | *P. f.* |
| MTIP-MyoA complex | 2AUC | | *P. k.* |
| Nucleoside diphosphate kinase B | 1XIQ | | *P. f.* |
| Ornithine aminotransferase | 1Z7D | | *P. y.* |
| Oxoacyl-ACP reductase | 2C07 | | *P. f.* |
| Phosphatidylethanolamine- | 2GZQ | | *P. v.* |
| binding protein | | | |
| Phosphoglycerate kinase | 1LTK | | *P. f.* |
| Phosphoglycerate mutase | 1XQ9 | SCN | *P. f.* |
| Plasmepsin | 1LS5, 1SME, 1ME6, 1LF2, 1LF4, 1PFZ, 1LEE, 1LF3, 1M43, 1MIQ, 2ANL, 1QS8, 1W6H, 1W6I, 2BJU, 1XE5, 1XE6, 1XDH | Peptstatin A, Inhibitor Rs370, Statine analogue, Inhibitor Rs367, Inhibitor Eh58, JE2 | *P. f.*, *P. v.*, *P. m.* |
| *P. f.* gamete antigen 27/25 | 1N81 | | *P. f.* |
| Purine nucleoside phosphorylase | 2BSX | Inosine | *P. f.* |
| Putative adenosine deaminase | 2AMX | | *P. y.* |
| Putative deoxyribose-phosphate aldolase | 2A4A | | *P. y.* |
| Putative, dim1 protein homolog | 2AV4 | | *P. y.* |
| putative FK506-binding protein PFL2275c | 2FBN | | *P. f.* |
| Putative formylmethionine deformylase | 1JYM, 1RQC, 1RL4 | BRR, BL5 | *P. f.* |
| putative HAD/COF-like hydrolase | 2B30 | | *P. v.* |
| Putative, heat shock protein | 1Y6Z | | *P. f.* |
| Putative, histamine-releasing factor | 1TXJ | | *P. k.* |
| Putative, orotidine-monophosphate-decarboxylase | 2AQW, 2FDS, 2GUU, 2FFC, UP6 | Sulphate, SeMethionine, Uridine-5'-monophosphate | *P. y.*, *P. b.*, *P. v.* |
| Putative uridine phosphorylase | 1Q1G, 1SQ6, 1NW4 | 5'- Methylthio-Immucillin-H, Immh | *P. f.* |
| Putative, vacuolar protein sorting 29 | 2BDD | | *P. y.* |
| Pvs25 | 1Z1Y, 1Z27 | | *P. v.* |
| Rab6 | 1D5C | GDP | *P. f.* |
| Ribose 5-phosphate isomerase | 2F8M | Phosphate | *P. f.* |
| Ribulose 5-phosphate 3-Epimerase | 1TQX | | *P. f.* |
| Spermidine synthase | 2FEG | MTA | *P. f.* |
| Thioredoxin | 1SYR, 2AV4 | | *P. f.*, *P. y.* |
| Thioredoxin peroxidase I | 2H01 | | *P. y.* |
| Triose Phosphate Isomerase | 1LYX, 1LZO, 1M7O, 1O5X, 1VGA, 1WOA, 1WOB, 1YDV, 1M7P | 2-Phosphoglycolic acid, Glycerol-3-Phosphate, 2-Phosphoglycolic acid, 3-Phosphoglyceric acid, 2-phosphoglycerate | *P. f.* |
| Ubiquitin conjugating enzyme E2 | 2H2Y, 2FO3 | | *P. f.*, *P. v.* |

Table 2